\documentclass[preprints,article,accept,moreauthors,pdftex]{Definitions/mdpi} 
\usepackage{amsfonts}
\usepackage{amssymb}
\usepackage{physics}
\usepackage{mathtools}
\usepackage{textcomp}
\usepackage{siunitx}
\usepackage{empheq}
\usepackage{bbold}
\usepackage{dcolumn}
\usepackage{bm}
\usepackage{wrapfig}
\usepackage{lipsum}

\firstpage{1} 
\pubvolume{1}
\issuenum{1}
\articlenumber{0}
\pubyear{2021}
\copyrightyear{2021}
\datereceived{26 October 2021}   
\dateaccepted{14 December 2021} 
\datepublished{{date}} 
\hreflink{https://doi.org/} 

\pdfoutput=1

\Title{Towards Quantum Simulation of Black Holes in a dc-SQUID~Array}
\TitleCitation{Towards Quantum Simulation of Black Holes in a dc-SQUID Array}
 
\Author{{Ad}ri\'an Terrones $^{1}$\orcidA{} and  Carlos Sab\'in $^{2,}$*\orcidB{}}

\AuthorNames{Adrián Terrones and Carlos Sabín}
\AuthorCitation{Terrones, A.; Sabín, C.}
\address{%
$^{1}$ \quad Departamento de F\'isica Te\'orica, Universidad Complutense de Madrid, Plaza de Ciencias 1, 28040 Madrid, Spain; adriterr@ucm.es\\
$^{2}$ \quad Departamento de F\'isica Te\'orica, Universidad Aut\'onoma de Madrid, 28049 Madrid, Spain}
\corres{Correspondence: carlos.sabin@uam.es}

\abstract{We propose quantum simulations of 1~+~1D radial sections of different black hole spacetimes (Schwarzschild, Reissner--Nordstr\o{}m, Kerr and Kerr--Newman), by means of a dc-SQUID array embedded on an open transmission line. This was achieved by reproducing the spatiotemporal dependence of 1~+~1D sections of the spacetime metric with the propagation speed of the electromagnetic field in the simulator, which can be modulated by an external magnetic flux. We show that the generation of event horizons---and therefore Hawking radiation---in the simulator could be achieved for non-rotating black holes, although we discuss limitations related to fluctuations of the quantum phase. In the case of rotating black holes, it seems that the simulation of ergospheres is beyond reach.}


\begin{document}



\section{Introduction}
Black holes are some of the most unique and fascinating objects in the universe, but a full understanding of them remains elusive, mostly due to their extreme nature and the lack of experimental support to guide and test our theoretical progress. In general, we cannot probe and take measurements of black holes on site. All we can do is observe from afar just some of their phenomena, and most of them are indirect events. However, the most interesting feature of a black hole is its event horizon (and its ergosphere in the case of rotating black holes), especially concerning how quantum field theory (QFT) processes operate in such particular gravitational backgrounds.

However, from a theoretical point of view, we have not yet found the way to make general relativity (GR) compatible with QFT~\cite{why_quantum_gravity_Woodard_2009}. This might also be related to other paradigmatic problems of GR, such as the existence of closed timelike curves (CTCs), which are allowed by GR and would break the principle of causality by allowing time travel. To fix this, the Chronology Protection Conjecture was proposed by Hawking~\cite{chronology_protection_PhysRevD.46.603}, by which quantum gravitational effects could prevent CTCs from existing, thereby preventing causal violations. Thus, we would need a full quantum theory of gravity to properly understand processes in the proximity of a black hole at a quantum level.

 How to fit gravity into QFT is a very active subject that has concerned theoretical physicists for many years, to this day. While QFT is a purely quantum theory, Einstein's general relativity is built completely from classical physics foundations. Hence, at very short lengths and times and in the vicinity of an extreme gravitational field, quantum fluctuations of spacetime are expected from a QFT point of view, but are incompatible with the classical behaviour of GR. The first approach to solve this issue is to quantize GR. However it is well-known that this approach produces a nonrenormalizable theory. Still, much progress has been made over the years to build a theory for \textls[-15]{quantum gravity, with the most prominent approaches being superstring theory and loop quantum gravity---}although they are still unsatisfactory, as they are far from making any measurable predictions.

To make some phenomenological progress, we can take a more modest path, where spacetime is kept as a classical background and is coupled to QFT, the same way a light--matter interaction can be approximated as a classical electromagnetic field coupled to a quantum system; this semiclassical approximation is known as quantum field theory in curved spacetime (QFTCS). This approach allows us to make use of the well-known perturbation theory, and extract useful calculations and predictions in accordance with the still not fully known theory of quantum gravity, which, just like many quantum electrodynamics results, can be obtained from semiclassical electromagnetic theory. The assumption of treating a gravitational field as a small perturbation breaks down when the length and time scales fall below the order of the Planck length $l_P=(G\hbar / c^3)^{1/2}=1.616 \times 10^{-33}~\text{cm}$ and Planck time $t_P=(G\hbar / c^5)^{1/2}=5.39 \times 10^{-44} \text{cm}$, as higher orders of the expansion become comparable with the first order, since the Planck length acts as the coupling constant in this approximation~\cite{birrell_davies_1982}. Moreover, we know gravity couples equally with all kinds of energy---including gravitational energy. Thus, a hypothetical graviton would couple to itself as much as other particles---for example, photons. Therefore, it is impossible to completely ignore the intrinsic nonlinearity of quantum gravity at any scale~\cite{birrell_davies_1982}.
Despite this issue, one can hope to obtain some useful results from the first order of expansion, except for very extreme regimes, such as those close to a microscopic black hole or in the first moments after the big bang~\cite{birrell_davies_1982}. In fact, great progress has been made in this field, which nowadays possesses a very solid foundation. The most famous application of QFTCS is probably black hole thermodynamics, which includes Hawking radiation~\cite{hawking_radiation_Hawking1975}, and has also led to the formulation of the holographic principle~\cite{holografic_principle_RevModPhys.74.825}. Other examples include the Unruh effect~\cite{unruh_effect_PhysRevD.14.870} or Hawking's aforementioned chronology protection conjecture.

All this progress is very good news from a theoretical point of view, as it seems to let us take a step towards quantum gravity. However, it does not solve the issue of the absence of experimental evidence presented earlier. In fact, predictions made from QFTCS are extremely hard to check empirically. The lack of experiments to support the current theoretical predictions---and boost more advanced theories and possible new physics---encourages us to make use of simulations, a very useful resource for research at our disposal. Simulations let us explore the physics of unreachable systems or hard to observe phenomena, but usually at the cost of having limitations of applicability. For instance, classical simulators have been proposed to reproduce systems in curved spacetimes, such as simulations of wormholes with electromagnetic metamaterials~\cite{EM_wormholes_PhysRevLett.99.183901}, water~\cite{water_wormholes_PhysRevD.93.084032} or magnetic metamaterials~\cite{magnetic_wormhole}; superluminal motion~\cite{superluminal_motion}; or the event horizon of a white hole~\cite{event_horizon}. However, to search for situations in which quantum fields are involved, we are compelled to turn instead to quantum simulators, where quantum gravitational effects could, in principle, arise. Some proposals for quantum simulations include traversable wormholes~\cite{bose_wormhole_PhysRevD.97.044045, squid_wormhole_PhysRevD.94.081501}, and searching for a chronology protection mechanism~\cite{CTCs_Mart_n_V_zquez_2020} or other exotic spacetimes, such as warp drives, G\"odel spacetimes or extreme Kerr black holes~\cite{exotic_spacetimes_Sab_n_2018}. 
These simulations can be performed in very different quantum setups, depending on the simulated system or what is expected to be achieved. For example, the previously cited simulations were proposed for a Bose--Einstein condensate~\cite{bose_wormhole_PhysRevD.97.044045} or a dc-SQUID array~\cite{squid_wormhole_PhysRevD.94.081501, exotic_spacetimes_Sab_n_2018, CTCs_Mart_n_V_zquez_2020}. A simulation of Hawking radiation has been successfully achieved in a laboratory with Bose--Einstein condensates~\cite{steinhauer}, and theoretical proposals with superconducting circuits can be found in the literature~\cite{quantum_vacuum_hawking_radiation_RevModPhys.84.1,PhysRevLett.103.087004}.

In this work, we followed the procedure of~\cite{exotic_spacetimes_Sab_n_2018}, which proposes the simulation of 1~+~1 dimensional sections of exotic spacetimes using a superconducting circuit consisting of a dc-SQUID array embedded in an open microwave transmission line~\cite{dynamical_casimir_effect,squid_array_Haviland2000,squid_array_Watanabe_PhysRevB.67.094505,squid_array_ergul_PhysRevB.88.104501}. We took this approach to propose the simulation of several black hole spacetimes, in order of complexity: the Schwarzschild black hole (chargeless and non-rotating), the Reissner--Nordstr\o m black hole (charged and non-rotating), the Kerr black hole (chargeless and rotating) and the Kerr--Newman black hole (charged and rotating). The objectives are to simulate radial sections of the spacetimes for both the interior and exterior of the black hole, generate event horizons and ergosurfaces and analyse possible behaviours close to the horizons that could be observed in an experiment, for all the spectrum of parameters $Q$ (charge) and $S$ (angular momentum), and radial directions given by the polar angle $\theta$.
We found that event horizons and Hawking radiation could, in principle, be simulated in our model for non-rotating black holes, although we discuss possible limitations. In the case of rotating black holes, our techniques seem not able to tackle the quantum simulation of the ergosphere. 
 
\section{Model and Results}

\subsection{Simulator}

A superconducting quantum interference device (SQUID) consists of two Josephson junctions (JJs) connected in parallel to a superconducting circuit. A Josephson junction is, in turn, a device formed by two superconducting leads coupled by a weak link, made by a thin insulating barrier, which allows one to implement the Josephson effect~\cite{JOSEPHSON1962251} in the superconducting circuit. In practice, if the area of the SQUID is sufficiently small, its self-inductance can be neglected, and if the two JJs possess identical critical currents, the SQUID can be viewed as a single JJ whose inductance can be controlled with an external magnetic field (see~\cite{simoen_2015,jjs_squids_You2011} for more details about JJs and SQUIDs). Hence, in a dc-SQUID array embedded in an open transmission line, the effective propagation speed of the electromagnetic field can be spatially and temporally modulated through the inductance of the circuit, which depends on the external magnetic flux passing through it, because
\begin{equation}
    c=\frac{1}{\sqrt{CL}}
\end{equation}
is the effective speed of light through the array, where $C$ and $L$ are the capacitance and inductance per unit length of the array, respectively, with~\cite{simoen_2015}
\begin{equation}
\label{eq:inductance}
    L_s (\phi_{\text{ext}}) = \frac{\phi_0}{4\pi I_c \abs{\cos{\frac{\pi \phi_{\text{ext}}}{\phi_0}}} \cos \psi}
\end{equation}
being the inductance of a single SQUID, where $I_c$ is its critical current, $\phi_0=h/2e$ is the magnetic flux quantum, $\phi_{\text{ext}}$ is the external magnetic flux and $\psi$ is the SQUID phase difference, which gives rise to nonlinearities. Therefore, we will work on the weak signal limit linear regime, where we assume $\cos\psi\simeq 1$~\cite{squid_wormhole_PhysRevD.94.081501, CTCs_Mart_n_V_zquez_2020,exotic_spacetimes_Sab_n_2018}. When no flux is applied, the effective speed of propagation all along the array is
\begin{equation}
    c_0\equiv c(\phi_{\text{ext}}=0)=\frac{\epsilon}{\sqrt{C_s L_s(\phi_{\text{ext}}=0)}}=\epsilon \sqrt{\frac{4\pi I_c}{\phi_0 C_s}} \,,
\end{equation}
where $\epsilon$ is the length of the SQUID. Then,
\begin{equation}
    c^2(\phi_{\text{ext}})=c_0^2 \abs{\cos{\frac{\pi \phi_{\text{ext}}}{\phi_0}}}
\end{equation}
gives us the speed of light as a function of the external magnetic flux. In turn, the flux can have spatial and temporal dependence, so we can split it into constant (DC) and variable (AC) contributions:
\begin{equation}
    \phi_{\text{ext}}(r,t) = \phi_{\text{ext}}^{\text{DC}} + \phi_{\text{ext}}^{\text{AC}}(r,t) \,.
\end{equation}

We can use the DC contribution to reduce the effective speed of light all along the array~\cite{exotic_spacetimes_Sab_n_2018}, so that we can define
\begin{equation}
    \tilde{c}_0^2 \equiv c^2(\phi_{\text{ext}}^{\text{AC}}=0) = c_0^2 \abs{\cos{\frac{\pi \phi_{\text{ext}}^{\text{DC}}}{\phi_0}}} \,,
\end{equation}
 and therefore,
\begin{equation}
\label{eq:total_speed_light_squid}
    c^2(\phi_{\text{ext}}) = \tilde{c}_0^2(\phi_{\text{ext}}^{\text{DC}}) \tilde{c}^2(\phi_{\text{ext}}) \,,
\end{equation}
where
\begin{equation}
\label{eq:speed_light_squid}
    \tilde{c}^2(\phi_{\text{ext}}) = \abs{\sec{\frac{\pi \phi_{\text{ext}}^{\text{DC}}}{\phi_0}}} \abs{\cos{\frac{\pi \phi_{\text{ext}}}{\phi_0}}} \,.
\end{equation}

If we keep fluxes in the interval $\pi \phi_{\text{ext}}/\phi_0 \in [-\pi/2,\pi/2]$, we can get rid of the absolute values and invert Equation (\ref{eq:speed_light_squid}) to obtain the magnetic flux necessary to induce a particular speed of light (with respect to the effective speed of light $\tilde{c}_0$ without external bias) through the array:
\begin{equation}
    \label{eq:flux_effective_speed}
    \frac{\pi\phi_{\text{ext}}^{\text{AC}}}{\phi_0} = \arccos \left[ \cos \left( \frac{\pi\phi_{\text{ext}}^{\text{DC}}}{\phi_0} \right) \tilde{c}^2 \right] - \frac{\pi\phi_{\text{ext}}^{\text{DC}}}{\phi_0}\,,
\end{equation}
which, in general, can depend on space and time.

The objective now is to obtain the 1~+~1D sections of the spacetimes produced by the black holes. Then, we can compare them 
 with Equation~(\ref{eq:total_speed_light_squid}), where $\tilde{c}_0$ is analogous to the speed of light in vacuum, and which, in fact, can be set to be significantly smaller to the speed of light in vacuum by choosing a suitable value of $\phi_{\text{ext}}^{\text{DC}}$. This way, we can indirectly simulate the 1~+~1D section of the spacetime through the speed of propagation of the electromagnetic field. In principle, there could be other ways to achieve a particular spatiotemporal dependence of the speed of propagation, even with superconducting setups~\cite{PhysRevB.77.144507}.

\subsection{Black Holes}

The metric of a general black hole with mass M, charge Q and angular momentum S is given by the Kerr--Newman metric in Boyer--Lindquist coordinates (and natural units $c=G=1$), with line element~\cite{Misner:1973prb}
\begin{equation}
\label{kn_metric}
    ds^2 = -\frac{\Delta}{\rho^2}\left[ dt-a\sin^2\theta \,d\phi \right]^2 + \frac{\sin^2\theta}{\rho^2}\left[ \left(r^2+a^2\right) d\phi - a\,dt \right]^2 + \frac{\rho^2}{\Delta}dr^2 + \rho^2 d\theta^2 \,,
\end{equation}
where
\begin{equation}
\label{kn_parameters}
    a = \frac{S}{M}\,, \quad \Delta = r^2-2Mr+a^2+Q^2\,, \quad \rho^2 = r^2+a^2\cos^2\theta \,.
\end{equation}

For convenience, we define the following dimensionless parameters:
\begin{equation}
    \xi\equiv\frac{r}{M}\,,\quad \tau\equiv \frac{t}{M}\,, \quad A\equiv \frac{Q}{M}\,, \quad B\equiv\frac{S}{M^2}\,, \quad C\equiv\cos \theta \,.
\end{equation}

Therefore,
\begin{equation}
    a = MB\,, \quad \Delta = M^2\left(\xi^2-2\xi+A^2+B^2\right)\,, \quad \rho^2 = M^2(\xi^2+B^2C^2) \,,
\end{equation}
so that we can write
\begin{equation}
\begin{aligned}
    ds^2 = M^2 \Bigg\{ &-\frac{\xi^2-2\xi+A^2+B^2}{\xi^2+B^2C^2} \Big[ d\tau - B(1-C^2)d\phi \Big]^2 + \\
    &+ \frac{1-C^2}{\xi^2+B^2C^2} \Big[ \left( \xi^2 + B^2 \right) d\phi - B d\tau \Big]^2 + \\
    &+ \frac{\xi^2+B^2C^2}{\xi^2-2\xi+A^2+B^2} d\xi^2 + \left( \xi^2+B^2C^2 \right) d\theta^2 \Bigg\} \,.
\end{aligned}
\end{equation}

Note that with this choice, the parameter $C^2=\cos^2\theta$ is restricted to the values $0\leq C^2\leq 1$. Additionally, the three externally observable parameters of a black hole must satisfy the inequality $Q^2 + \frac{S^2}{M^2} \leq M^2$~\cite{Misner:1973prb}. Therefore, $A$, $B$ and $C$ must fulfil the constraints
\begin{equation}
\label{eq:constraints}
    \begin{dcases}
    A^2 + B^2 \leq 1 \\
    B^2C^2 \leq B^2
    \end{dcases} \,,
\end{equation}
which will be relevant in the subsequent analysis. Black holes in the limit $A^2 + B^2 = 1$ are known as extremal black holes.

The main feature of a black hole is the event horizon, i.e., the boundary through which no particle or signal can escape from the interior to the exterior. Matter and light can fall inside through it, but in Boyer--Lindquist coordinates, which correspond to an observer at infinity, it will take an infinite time for an infalling particle or signal to reach the horizon. To remove this singularity, other coordinates can be chosen, such as the ones attached to an infalling observer, who would fall through the horizon in a finite, proper amount of time~\cite{Misner:1973prb}. We take this into account in our subsequent discussions.

Black holes with nonzero angular momentum possess another interesting feature, which is the dragging of inertial frames. This means that static observers in a reference frame close to the black hole have nonzero angular momentum according to observers at infinity~\cite{Misner:1973prb}. This effect becomes stronger when we approach the black hole, until the dragging is so intense that no observer can be static according to another at infinity. The region where this occurs is called the ergosphere, and is limited by the static limit or ergosurface on the outside and the event horizon on the inside.

Another thing we will have to take into account is that the metric in Equation (\ref{kn_metric}) can only be related to a physical black hole at and outside the event horizon, but not inside it~\cite{Misner:1973prb}. We will therefore focus the analysis on the regions of the simulations close to the event horizon and outside it. However, we still comment on the inner parts of the simulated black holes, as they could be interesting cases of exotic spacetimes, even if they do not represent actual black holes.

It can be shown~\cite{Misner:1973prb} that a black hole has event horizons at
\begin{equation}
    r = r_\pm = M \pm \sqrt{M^2 - Q^2 - a^2}
\end{equation}
in Boyer--Lindquist coordinates. Using our notation, the event horizons are located at
\begin{equation}
\label{eq:event_horizon}
    \xi = \xi_\pm = 1 \pm \sqrt{1 - A^2 - B^2} \,,
\end{equation}
where $\xi_+ \ge \xi_-$. Therefore, the horizons at $\xi_+$ and $\xi_-$ are known as outer and inner event horizons, respectively (the outer event horizon is the one we consider the \textit{true} event horizon, which defines the boundary between the interior and the exterior of the black hole; the same can be said about the outer ergosurface later on). These two horizons coincide only when $A^2+B^2 = 1$ at $\xi=\xi_+=\xi_-=1$, which corresponds to a extremal black hole. When $A=B=0$ (Schwarzschild geometry), the inner event horizon is at $\xi=\xi_-=0$, which corresponds to a physical singularity of the spacetime. Hence, the Schwarzschild black hole presents only one event horizon.

On the other hand, the surfaces known as static limits or ergosurfaces that set the boundaries of the ergosphere are located at~\cite{Misner:1973prb}
\begin{equation}
    r = r_\pm^E = M \pm \sqrt{M^2 - Q^2 - a^2\cos^2\theta},
\end{equation}
which correspond to
\begin{equation}
\label{eq:ergosurface}
    \xi = \xi_\pm^E = 1 \pm \sqrt{1 - A^2 - B^2C^2} \,.
\end{equation}

As $\xi_+^E \ge \xi_-^E$, $\xi_+^E$ is known as the outer static limit and $\xi_-^E$ as the inner static limit. Taking into account the restrictions in Equation (\ref{eq:constraints}), we can see that
\begin{equation}
    2 \ge \xi_+^E \ge \xi_+ \ge 1 \ge \xi_- \ge \xi_-^E \ge 0 \,.
\end{equation}

The region between the outer ergosurface and the outer event horizon, where no observer cannot ever be at rest, is what we call the ergosphere. At the rotation axis of the black hole (that is, when $\theta=0,\pi$), the outer static limit and event horizon coincide, as do their inner counterparts. When $B=0$, $\xi_+^E = \xi_+$ (and $\xi_-^E = \xi_-$) for all $\theta$, and therefore no ergosphere is formed at spinless black holes.

By reducing the metric to 1~+~1D for the variables $\{\tau,\xi\}$ ($\{t,r\}$), we obtain
\begin{equation}
    ds_{\text{1~+~1D}}^2 = M^2\Bigg( -\frac{\xi^2-2\xi+A^2+B^2C^2}{\xi^2+B^2C^2} d\tau^2 + \frac{\xi^2+B^2C^2}{\xi^2-2\xi+A^2+B^2} d\xi^2 \Bigg) \,.
\end{equation}

Now, notice that, following~\cite{exotic_spacetimes_Sab_n_2018}, we can exploit the conformal invariance of the Klein--Gordon equation to write this metric in the form $ds_{\text{1~+~1D}}^2 = \tilde{c}^2(\xi) d\tau^2 + d\xi^2$, where $\tilde{c}^2(\xi)$~is:
\begin{equation}
    \tilde{c}^2(\xi) = \frac{\left(\xi^2-2\xi+A^2+B^2\right)\left(\xi^2-2\xi+A^2+B^2C^2\right)}{\left(\xi^2+B^2C^2\right)^2},
\end{equation}
and by setting $\phi_{\text{ext}}^{\text{DC}}=0$ in Equation (\ref{eq:flux_effective_speed}) we find
\begin{equation}
    \frac{\pi\phi_{\text{ext}}(\xi)}{\phi_0}=\arccos{\left[ \frac{\left(\xi^2-2\xi+A^2+B^2\right)\left(\xi^2-2\xi+A^2+B^2C^2\right)}{\left(\xi^2+B^2C^2\right)^2} \right]} \,.
\end{equation}




\section{Results}

\subsection{Schwarzschild}

In the limit $Q=S=0$ (non-charged static black hole), we have
\begin{equation}
    A=0\,, \quad B=0 \,.
\end{equation}

Therefore,
\begin{gather}
    ds_{\text{1~+~1D}}^2=M^2\left[-\left( 1-\frac{2}{\xi} \right)d\tau^2+\frac{1}{1-\frac{2}{\xi}}d\xi^2\right] \,, \\
    \tilde{c}^2(\xi)=\left( 1-\frac{2}{\xi} \right)^2 \,, \\
    \frac{\pi\phi_{\text{ext}}(\xi)}{\phi_0}=\arccos{\left[ \left( 1-\frac{2}{\xi} \right)^2 \right]} \,.
\end{gather}

From Equation (\ref{eq:event_horizon}), we expect to see only one event horizon at $\xi=\xi_+=2$, as discussed in the previous section; and no ergosphere, given that the Schwarzschild black hole has no angular momentum.

\begin{figure}[H]
\includegraphics[width=10.5 cm]{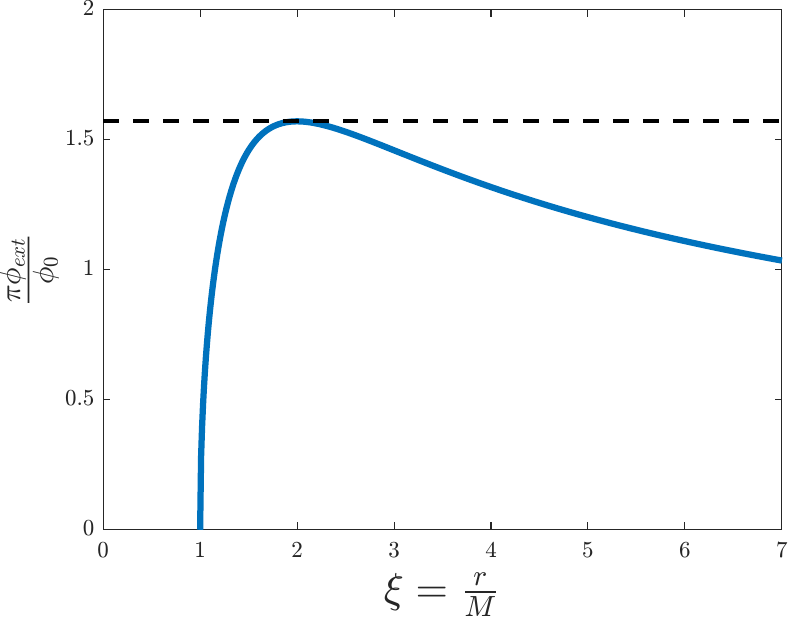}
\caption{External magnetic flux $\pi\phi_{\text{ext}}/\phi_0$ versus the adimensional distance $\xi=r/M$ for the simulation of a Schwarzschild black hole. The dashed line represents the threshold value \mbox{$\pi\phi_{\text{ext}}/\phi_0=\pi/2$}.}
\label{fig:sch}
\end{figure}

We can see in Figure \ref{fig:sch} that the flux profile $\pi \phi_{\text{{ext}}}/\phi_0$ is well defined for $\xi \geq 1$, but not for the interval $0 \leq \xi < 1$. It presents a maximum at $\xi=2$, when it reaches the threshold value of $\pi/2$, whereas for big values of $\xi$ the flux tends to zero. This behaviour is just what we expected, as the maximum coincides with the event horizon, because $\pi \phi_{\text{{ext}}}/\phi_0=\pi/2$ corresponds to effective light speed $\tilde{c}^2 = 0$; and $\pi \phi_{\text{{ext}}}/\phi_0=0$ at infinity corresponds to $\tilde{c}^2 = 1$, as it should be for an asymptotically flat spacetime.

\begin{figure}[H]
\includegraphics[width=10.5 cm]{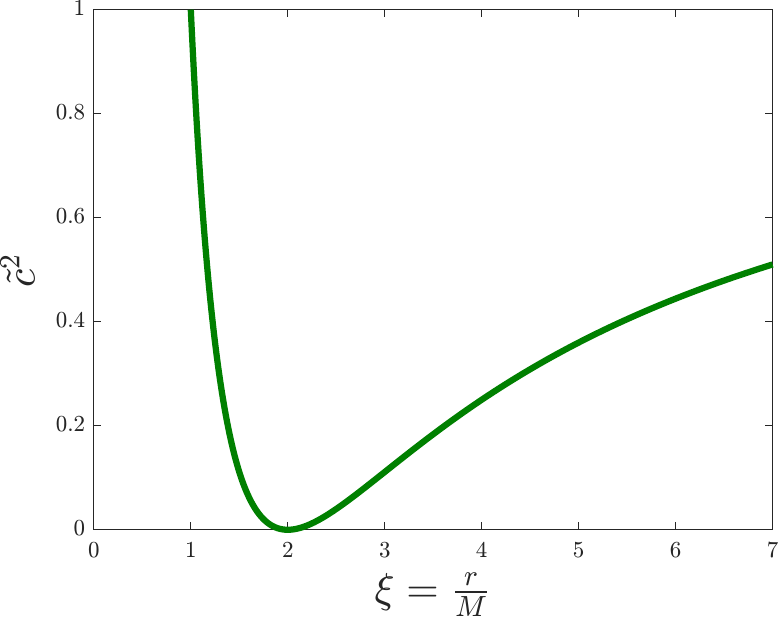}
\caption{Effective speed of light squared $\tilde{c}^2$ versus the adimensional distance $\xi=r/M$ for the simulation of a Schwarzschild black hole.}
\label{fig:sch_c}
\end{figure}

In fact, one can think of the flux profile as something like a potential barrier for the electromagnetic field: as the flux increases, the effective speed of light decreases, until it reaches the limit value of $\pi/2$ and light stops, as we can see by comparing Figure \ref{fig:sch_c} to Figure \ref{fig:sch}. This means that a signal sent from inside the black hole cannot ever reach the exterior, just as the definition of an event horizon implies. However, this seems to suggest that a signal sent from the exterior cannot reach the interior either. However, we must remember that the coordinates we are working with are defined for an observer at infinity, so any matter or light falling towards the black hole will take infinite time to cross the event horizon, whereas for the coordinate system of an infalling observer, the same process will take a finite, proper amount of time~\cite{Misner:1973prb}. Therefore, a signal can penetrate the event horizon, even if an observer far away cannot ever see it.

Unfortunately, the very event horizon that we are looking to simulate presents a technical issue. When the flux is close to the limit value $\pi \phi_{\text{{ext}}}/\phi_0=\pi/2$, quantum fluctuations appear due to the very high impedance of the array. If a large number of SQUIDs are close to the limit, this could lead to fluctuations in the superconducting phase $\psi$ becoming dominant, breaking down our approximation $\cos \psi \simeq 1$ and eventually preventing the array for being in the superconducting phase~\cite{squid_wormhole_PhysRevD.94.081501, exotic_spacetimes_Sab_n_2018,squid_array_Haviland2000}. Therefore, we should try to avoid this behaviour by keeping the least possible SQUIDs close to the limit value of the flux, ideally to just a single SQUID, by choosing values for $M$ so that the critical region for $r$ in the simulator is small. This makes even more sense if we realise that in our theoretical framework of general relativity, spacetime is continuous, so the event horizon in a 1~+~1D section is just a point in space, whereas the simulation is made from a discrete number of SQUIDs. Hence, a single SQUID is the ideal minimum we can achieve to accommodate the event horizon and its surroundings.

It has been proposed~\cite{quantum_vacuum_hawking_radiation_RevModPhys.84.1} and demonstrated~\cite{dynamical_casimir_effect} that it is possible to produce photon pairs from the quantum vacuum in superconducting circuit devices,  for instance, in SQUID arrays. Some of those could be interpreted as Hawking radiation when an analogue event horizon is produced in the simulator. In fact, when a correlated pair of photons is created due to quantum fluctuations of the vacuum very close to the horizon, where one photon is emitted towards the interior of the black hole and the other to the exterior, we can measure radiation coming from the simulated black hole~\cite{quantum_vacuum_hawking_radiation_RevModPhys.84.1}. Therefore, if we can indeed produce an event horizon in our simulation, we should be able to observe Hawking radiation. However, as commented above, our current theoretical understanding of this system is limited to the linear regime, where the SQUID phase is taken to satisfy the approximation $\cos \psi \simeq 1$. In fact, experiments proposed and performed with SQUID arrays tend to work only with low magnetic fluxes. Hence, we would need to experimentally carry out the simulation in order to check if Hawking radiation is actually produced at the horizon, or if, on the contrary, the nonlinearities in that region prevent the event horizon from forming in the simulation. With this setup, both emitted correlated photons could be detected at each end of the array. To make sure that this radiation is indeed emitted at the event horizon and is not just thermal noise emission, the experiment should measure the radiation both outside and inside the horizon and check that both correlated photons reach the detectors through coincidence detection~\cite{quantum_vacuum_hawking_radiation_RevModPhys.84.1}.

\subsection{Reissner--Nordstr\o m}

The metric of a non-rotating charged black hole ($S=0$) satisfies
\begin{equation}
    B=0 \,.
\end{equation}

Therefore,
\begin{gather}
    ds_{\text{1~+~1D}}^2=M^2\Bigg[-\left( 1-\dfrac{2}{\xi}+\dfrac{A^2}{\xi^2} \right)d\tau^2 + \left(1-\dfrac{2}{\xi}+\dfrac{A^2}{\xi^2}\right)^{-1}d\xi^2\Bigg] \,, \\
    \tilde{c}^2(\xi)=\left( 1-\dfrac{2}{\xi}+\dfrac{A^2}{\xi^2} \right)^2 \,, \\
    \frac{\pi\phi_{\text{ext}}(\xi)}{\phi_0}=\arccos{\left[ \left( 1-\dfrac{2}{\xi}+\dfrac{A^2}{\xi^2} \right)^2 \right]} \,.
\end{gather}

Looking at the preceding equations, depending on the value of $\abs{A}$, we expect to see four different behaviours for $\pi\phi_{\text{ext}}(\xi)/\phi_0$:
\begin{enumerate}
    \item If $A=0$:
\end{enumerate}
\begin{equation}
\begin{gathered}
    \xi \in \left[ 1 , +\infty \right] \,, \\
    \frac{\pi\phi_{\text{ext}}(\xi)}{\phi_0}=\frac{\pi}{2} \iff \xi=2 \,.
\end{gathered}
\end{equation}
\begin{enumerate}
\setcounter{enumi}{1}
    \item If $0<\abs{A}<\frac{1}{\sqrt{2}}$:
\end{enumerate}
\begin{equation}
\begin{gathered}
    \xi \in \left[ \frac{A^2}{2} , \frac{1}{2}\left( 1-\sqrt{1-2A^2} \right) \right] \cup \left[\frac{1}{2}\left( 1+\sqrt{1-2A^2} \right) , +\infty \right] \,, \\
    \frac{\pi\phi_{\text{ext}}(\xi)}{\phi_0}=\frac{\pi}{2} \iff \xi=1\pm \sqrt{1- A^2} \,.
\end{gathered}
\label{eq:if2}
\end{equation}
\begin{enumerate}
\setcounter{enumi}{2}
    \item If $\frac{1}{\sqrt{2}} \leq \abs{A} < 1$:
\end{enumerate}
\begin{equation}
\begin{gathered}
    \xi \in \left[ \frac{A^2}{2} , +\infty \right] \,, \\
    \frac{\pi\phi_{\text{ext}}(\xi)}{\phi_0}=\frac{\pi}{2} \iff \xi=1\pm \sqrt{1-A^2} \,.
\end{gathered}
\label{eq:if3}
\end{equation}
\begin{enumerate}
\setcounter{enumi}{3}
    \item If $\abs{A} = 1$:
\end{enumerate}
\begin{equation}
\begin{gathered}
    \xi \in \left[ \frac{1}{2} , +\infty \right] \,, \\
    \frac{\pi\phi_{\text{ext}}(\xi)}{\phi_0}=\frac{\pi}{2} \iff \xi=1 \,.
\end{gathered}
\end{equation}
The interval $\abs{A}>1$ is forbidden by the condition $A^2 + B^2 = A^2 \leq 1$. Hence, we must study the simulation for different values of the parameter $A$.

\begin{figure}[H]
\includegraphics[width=10.5 cm]{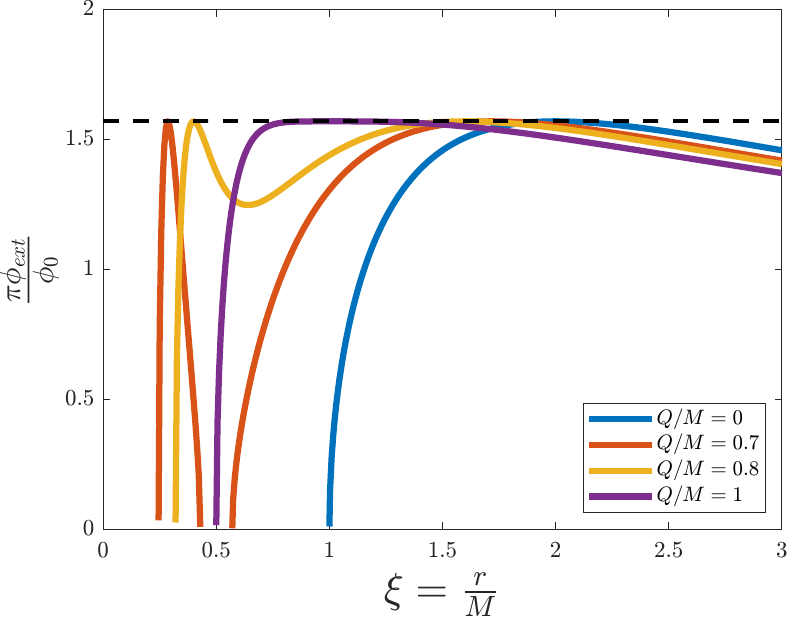}
\caption{External magnetic flux $\pi\phi_{\text{ext}}/\phi_0$ versus the adimensional distance $\xi=r/M$ for the simulation of a Reissner--Nordstr\o m black hole, for different values of $A=Q/M$. The dashed line represents the threshold value $\pi\phi_{\text{ext}}/\phi_0=\pi/2$.}
\label{fig:r_n}
\end{figure}

We can see in Figure \ref{fig:r_n} representations of four flux profiles for different values of $A=Q/M$. Again, no analogue of an ergosphere is present in any profile, as we expect for a non-rotating black hole. The blue line represents the limit of $Q=0$, which is equivalent to the Schwarzschild black hole, and yields the same flux as in Figure \ref{fig:sch}, as expected. The purple line corresponds to the extremal Reissner--Nordstr\o m black hole ($A=1$ or $Q=M$), which presents only one event horizon, as was discussed earlier for extremal black holes, at $\xi=1$. However, there is a big region where the flux is very close to its critical value, roughly between $\xi=0.7$ and $\xi=1.75$. To avoid problems in the simulation of an extremal Reissner--Nordstr\o m black hole, only large values for $M$ could be taken in order to minimise the critical region in the experiment. The red and yellow lines correspond to intermediate values of $A$. They present two event horizons (outer and inner) located in accordance with Equation (\ref{eq:event_horizon}) in $\xi=1\pm \sqrt{1-A^2}$. While the yellow profile is well defined from $\xi=A^2/2$ to infinity, the red one presents a region between the two horizons that cannot be simulated. These are the behaviours represented earlier in Equations (\ref{eq:if2}) and (\ref{eq:if3}); profiles with $A<1/\sqrt{2}\approx 0.7071$ present a discontinuity in the flux, whereas for $A\geq 1/\sqrt{2}$ the flux is continuous between the horizons.

The same can be said in this case as with the Schwarzschild black hole with respect to black hole thermal radiation. If we can keep the region with high magnetic flux in a small region of the array, 
ideally a single SQUID, in order to maintain the superconducting phase and low impedance throughout the circuit, we might be able to create an event horizon in the system, which could potentially generate Hawking radiation from quantum fluctuations of the vacuum. An experimental implementation of the simulation would be needed to verify this.

\subsection{Kerr}

A rotating non-charged black hole ($Q=0$) has
\begin{equation}
    A=0 \,.
\end{equation}

Therefore,
\begin{gather}
    ds_{\text{1~+~1D}}^2 = M^2\bigg( -\frac{\xi^2-2\xi+B^2C^2}{\xi^2+B^2C^2} d\tau^2 + \frac{\xi^2+B^2C^2}{\xi^2-2\xi+B^2} d\xi^2 \bigg) \,, \\
    \tilde{c}^2(\xi) = \frac{\left(\xi^2-2\xi+B^2\right)\left(\xi^2-2\xi+B^2C^2\right)}{\left(\xi^2+B^2C^2\right)^2} \,, \\
    \label{eq:kerr_flux}
    \frac{\pi\phi_{\text{ext}}(\xi)}{\phi_0}=\arccos{\left[ \frac{\left(\xi^2-2\xi+B^2\right)\left(\xi^2-2\xi+B^2C^2\right)}{\left(\xi^2+B^2C^2\right)^2} \right]} \,.
\end{gather}

We must now study the simulation for different sets of values of two parameters, $B$ and $C$. Note that $\theta$ ($C=\cos\theta$) here is a parameter, not a coordinate, which can take values in the interval $[0,\pi]$. However, due to the axial symmetry of this spacetime, we can restrict ourselves to $\theta\in[0,\pi/2]$, as the interval $[\pi/2,\pi]$ is equivalent. This is the same as taking $C$ between 1 and 0---again, because of the conditions in Equation (\ref{eq:constraints}), $B\leq 1$.

\begin{figure}[H]
\includegraphics[width=10.5 cm]{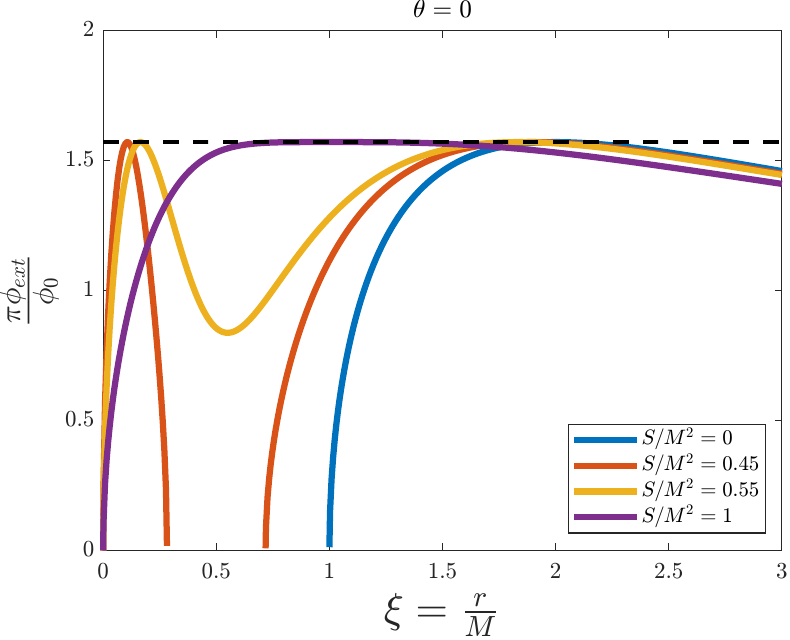}
\caption{External magnetic flux $\pi\phi_{\text{ext}}/\phi_0$ versus the adimensional distance $\xi=r/M$ for the simulation of a Kerr black hole with fixed $\theta=0$, and for different values of $B=S/M^2$. The dashed line represents the threshold value $\pi\phi_{\text{ext}}/\phi_0=\pi/2$.}
\label{fig:kr_0}
\end{figure}

We can see in Figure \ref{fig:kr_0} various magnetic flux profiles for the black hole's polar radial section ($\theta=0$). Even though this is a rotating black hole, the profiles are very similar to the ones for Reissner--Nordstr\o m, with no indication of an ergosphere. As commented earlier, this is because at the poles of the black hole the event horizons coincide with the static limits, so the ergosphere is not visible in this spacetime section. The only difference from the Reissner--Nordstr\o m black holes is that now we can simulate all the way down to $\xi=0$ (except when $B=0$), whereas in Reissner--Nordstr\o m black holes we are limited by $\xi=A^2/2$. The blue line represents the limit $B=0$, which is once again a Schwarzschild black hole. The purple line is now an extremal Kerr black hole ($B=1$ or $S=M^2$), which again has only one event horizon at $\xi=1$. This extremal black hole profile also presents a large problematic region, even larger than the Reissner--Nordstr\o m one, approximately between $\xi=0.5$ and $\xi=2$. This restricts the simulation even more to choosing only large values for the black hole's mass. For profiles with intermediate values of $B$ (red and yellow lines) we see the same behaviour as in the Reissner--Nordstr\o m black hole case, where we find two event horizons at $\xi=1\pm\sqrt{1-B^2C^2}$ and where profiles with $B$ below some value split between the inner and outer horizons. In general, for a Kerr black hole it is not possible to analytically calculate when this splitting occurs, except for this case ($\theta=0$), which occurs at $B=1/2$, as we can see with the red (below $B=1/2$) and yellow (above $B=1/2$) profiles.

\begin{figure}[H]
\includegraphics[width=10.5 cm]{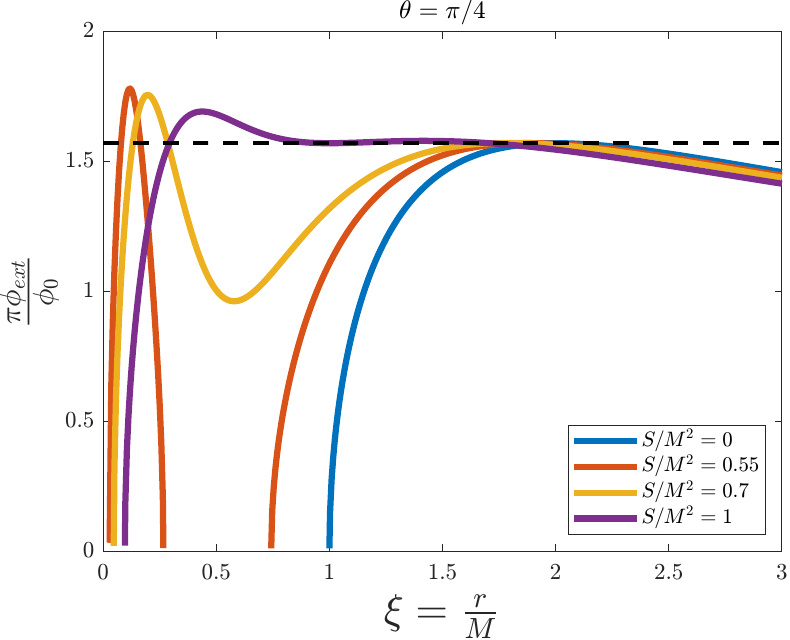}
\caption{External magnetic flux $\pi\phi_{\text{ext}}/\phi_0$ versus the adimensional distance $\xi=r/M$ for the simulation of a Kerr black hole with fixed $\theta=\pi/4$, and for different values of $B=S/M^2$. The dashed line represents the threshold value $\pi\phi_{\text{ext}}/\phi_0=\pi/2$.}
\label{fig:kr_pi4}
\end{figure}

\begin{figure}[H]
\includegraphics[width=10.5 cm]{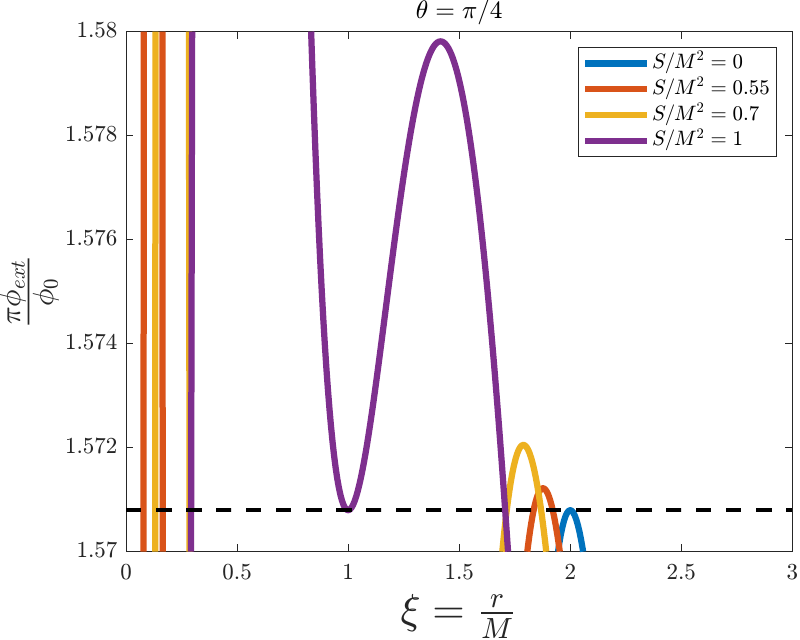}
\caption{A zoom-in of Figure \ref{fig:kr_pi4}, where the regions with $\pi\phi_{\text{ext}}/\phi_0$ above the threshold value $\pi/2$ (dashed line) can be better appreciated.}
\label{fig:kr_pi4_detail}
\end{figure}

In Figure \ref{fig:kr_pi4}, we have fixed $\theta=\pi/4$. Again, the limit $B=0$ is represented by the blue line and corresponds to the Schwarzschild black hole. However, for the rest of the profiles we can see a substantial change with respect to $\theta=0$. Now the red and yellow profiles take the threshold value $\pi \phi_{\text{{ext}}}/\phi_0=\pi/2$ four times, and the purple one three times, which correspond to the two event horizons and the two static limits or ergosurfaces (the purple profile is the Kerr extremal black hole, for which the inner and outer event horizons always coincide at $\xi=1$). For nonvanishing values of angular momentum, when coming from the right (large $\xi$) towards the centre of the black hole, the flux increases until it not only reaches $\pi \phi_{\text{{ext}}}/\phi_0=\pi/2$, but surpasses it (see in detail in Figure \ref{fig:kr_pi4_detail}). This is the outer ergosurface at $\xi = 1 + \sqrt{1 - B^2C^2}$. After that there is a region where the magnetic flux takes values above $\pi/2$ until it decreases and crosses the limit value again at $\xi=1 + \sqrt{1 - B^2}$, the outer event horizon. This region is what we call the ergosphere. After that we are inside the black hole and the same discussion applies for the inner event horizon and static limit. Again, a splitting occurs in the flux below a value of $B$ between $B=0.55$ and $B=0.7$.
\vspace{-9pt}

\begin{figure}[H]
\includegraphics[width=10.5 cm]{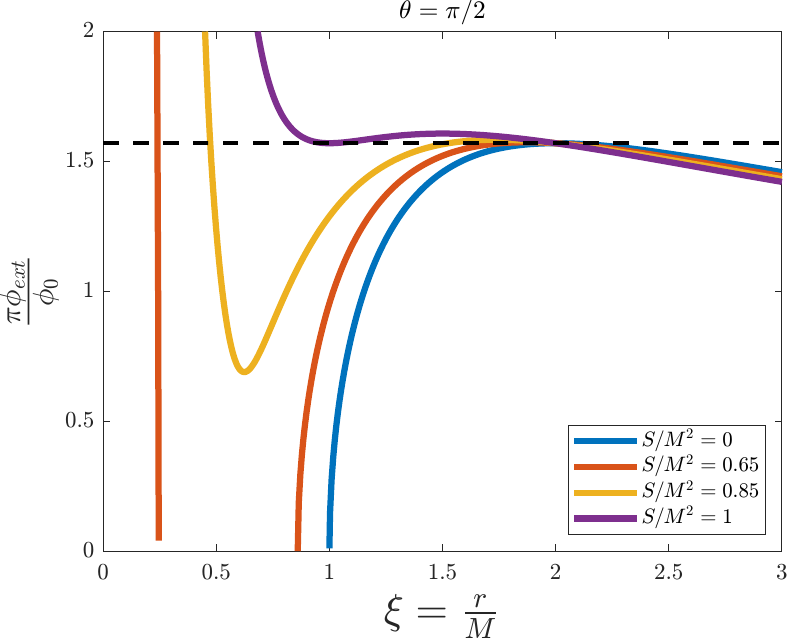}
\caption{External magnetic flux $\pi\phi_{\text{ext}}/\phi_0$ versus the adimensional distance $\xi=r/M$ for the simulation of a Kerr black hole with fixed $\theta=\pi/2$, and for different values of $B=S/M^2$. The dashed line represents the threshold value $\pi\phi_{\text{ext}}/\phi_0=\pi/2$.}
\label{fig:kr_pi2}
\end{figure}

In Figure \ref{fig:kr_pi2}, we have the radial section of spacetime at the equator of the black hole ($\theta=\pi/2$). The profiles of the flux here behave in a similar way as in Figure \ref{fig:kr_pi4}. Now the flux takes even greater values inside the ergosphere, and the flux splitting occurs at a higher value of $B$, between $B=0.65$ and $B=0.85$. The main difference is that the inner static limit is absent, as the flux diverges when it approaches small values of $\xi$ (except for the Schwarzschild limit $B=0$). Therefore, there is a polar angle $\theta$ between $\pi/4$ and $\pi/2$ for which we cannot simulate the inner static limit anymore. However, this is not a big issue, as we are mainly interested in what happens at the vicinities of the outer event horizon and static limit, and at the region between them, the ergosphere.

Focusing on the ergosphere for these figures, we can realize that a flux value over $\pi/2$ means that the square of the effective speed of light is negative in that region. Therefore, the effective speed of light inside the ergosphere must be imaginary, which makes no physical sense. In fact, note that with our theoretical frame we cannot consider values for the flux greater than $\pi/2$ in our system (see the discussion prior to Equation (\ref{eq:flux_effective_speed})), so  simulation of that region is not possible. Moreover, it seems that the static limit behaves the same way as an event horizon in our simulation, but theory tells us that it is not so: it is possible to escape and send signals from inside the static limit, provided that we have not crossed the event horizon. This issue comes not from the simulation, but actually from what we want to simulate, because our simulation does not cover the whole system, only radial sections of it. Additionally, by the definition of the static limit and the ergosphere, we know that inside it the effect of frame dragging is so intense that nothing, including light, can stay still with respect to objects at infinity, and must therefore rotate with the black hole~\cite{Misner:1973prb}. Hence, no light can travel through the ergosphere radially, and an imaginary effective speed of light mathematically represents that. That means that this kind of setup cannot be used to simulate an ergosphere, and therefore it is not suitable for simulating rotating black holes.

Of course, this also means that there is no hope for detecting any Hawking radiation from analogue rotating black holes simulated in this way. Even though we can generate an event horizon, only the region of spacetime directly inside it can be simulated, but not the region right outside it, which lies in the ergosphere. For this reason, the same can be said about the extraction of energy from the ergosphere outside the static limit. The spacetime outside the ergosphere is the only region of interest that can be simulated for a rotating black hole.

\subsection{Kerr--Newman}

For a general rotating and charged black hole, we have  \vspace{6pt}
\begin{gather}
    ds_{\text{1~+~1D}}^2 = M^2\bigg( -\frac{\xi^2-2\xi+A^2+B^2C^2}{\xi^2+B^2C^2} d\tau^2 + \frac{\xi^2+B^2C^2}{\xi^2-2\xi+A^2+B^2} d\xi^2 \bigg) \,, \\
    \tilde{c}^2(\xi) = \frac{\left(\xi^2-2\xi+A^2+B^2\right)\left(\xi^2-2\xi+A^2+B^2C^2\right)}{\left(\xi^2+B^2C^2\right)^2} \,, \\
    \frac{\pi\phi_{\text{ext}}(\xi)}{\phi_0}=\arccos{\left[ \frac{\left(\xi^2-2\xi+A^2+B^2\right)\left(\xi^2-2\xi+A^2+B^2C^2\right)}{\left(\xi^2+B^2C^2\right)^2} \right]} \,.
\end{gather}

Now we have to take into account all three parameters $A$, $B$ and $C$ (charge, angular momentum and polar angle) for the simulation. As with the Kerr black hole, it is sufficient to consider $C$ from 1 to 0 ($\theta\in[0,\pi/2]$) thanks to the axial symmetry of the spacetime. In this case we have to keep in mind the black hole condition set by Equation (\ref{eq:constraints}) for the set of parameters $\{A,B\}$. Therefore, by fixing $A$, $B$ is restricted to the interval $[0,\sqrt{1-A^2}]$.

\begin{figure}[H]
\includegraphics[width=10.5 cm]{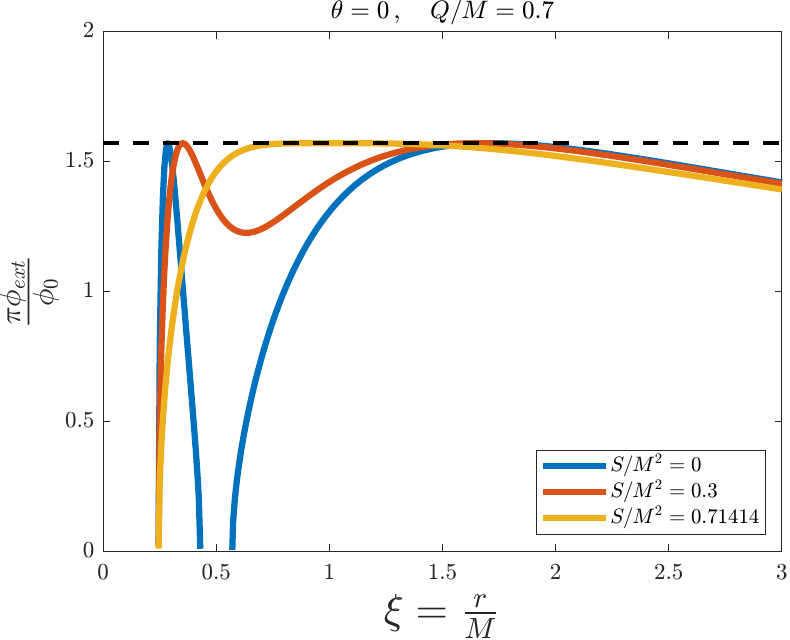}
\caption{External magnetic flux $\pi\phi_{\text{ext}}/\phi_0$ versus the adimensional distance $\xi=r/M$ for the simulation of a Kerr--Newman black hole with fixed $\theta=0$ and $A=Q/M=0.7$, and for different values of $B=S/M^2$. The dashed line represents the threshold value $\pi\phi_{\text{ext}}/\phi_0=\pi/2$.}
\label{fig:kn_0_0.7}
\end{figure}

In Figure \ref{fig:kn_0_0.7},  $C=1$ ($\theta=0$) and $A=0.7$. Thus, as with the Kerr black hole at the axis of rotation (see Figure \ref{fig:kr_0}), the ergosphere is not visible even with nonzero angular momentum (red and yellow lines). The blue line is the limit of no angular momentum, and therefore is a Reissner--Nordstr\o m black hole. It has $A=0.7<1/\sqrt{2}\approx 0.7071$, so splitting of the flux occurs between the inner and outer horizons (see discussion at the Reissner--Nordstr\o m subsection). The red profile with $B=0.3$ presents no splitting, so for some value of $B$ between 0 and 0.3 the spacetime is simulable between the horizons. Note that as in Reissner--Nordstr\o m black holes, the spacetime cannot be simulated for small values of $r$, as opposed to Kerr black holes, which are simulable down to $r=0$. Therefore, this seems to be a consequence of nonzero charge. The yellow profile corresponds to an extremal Kerr--Newman black hole ($B=\sqrt{1-A^2}$), with coincident event horizons at $\xi=1$. As the ergosphere is infinitely thin at this radial section, we can simulate it for all possible values of angular momentum and obtain event horizons, which could emit Hawking radiation. However, once again, the extremal black hole presents a large region of magnetic flux close to its threshold value. To avoid technical issues, only large values of M could be simulated in this case, in order to minimise the region of the simulator with high flux values.

\begin{figure}[H]
\includegraphics[width=10.5 cm]{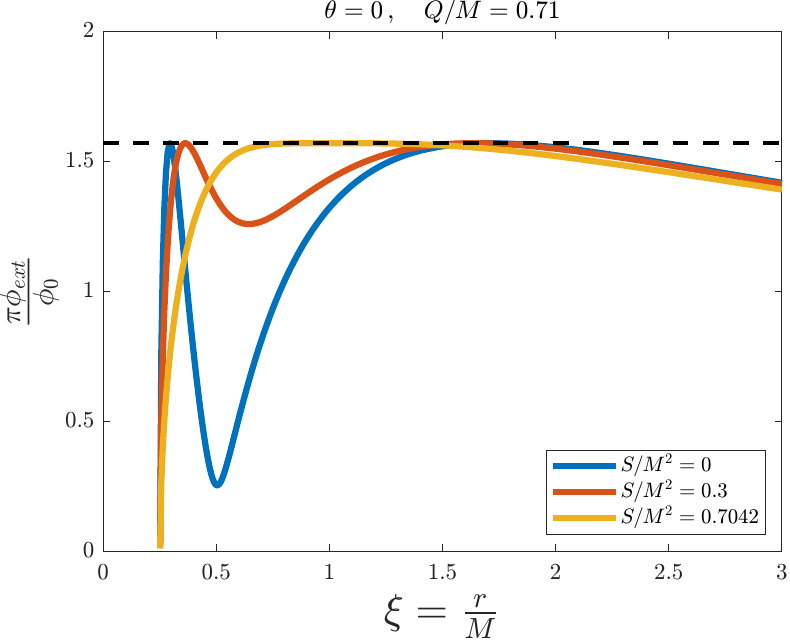}
\caption{External magnetic flux $\pi\phi_{\text{ext}}/\phi_0$ versus the adimensional distance $\xi=r/M$ for the simulation of a Kerr--Newman black hole with fixed $\theta=0$ and $A=Q/M=0.71$, and for different values of $B=S/M^2$. The dashed line represents the threshold value $\pi\phi_{\text{ext}}/\phi_0=\pi/2$.}
\label{fig:kn_0_0.71}
\end{figure}

For Figure \ref{fig:kn_0_0.71}, we kept $\theta=0$ but changed $A$ to a slightly higher value $A=0.71$. We can see no flux splitting in the Reissner--Nordstr\o m limit (blue line), because $A>1/\sqrt{2}$, and thus no splitting appears for any value of $B$. The rest of the profiles behave exactly the same way as in Figure \ref{fig:kn_0_0.7}, with the yellow line representing the extremal Kerr--Newman black hole for this value of $A$.

\begin{figure}[H]
\includegraphics[width=10.5 cm]{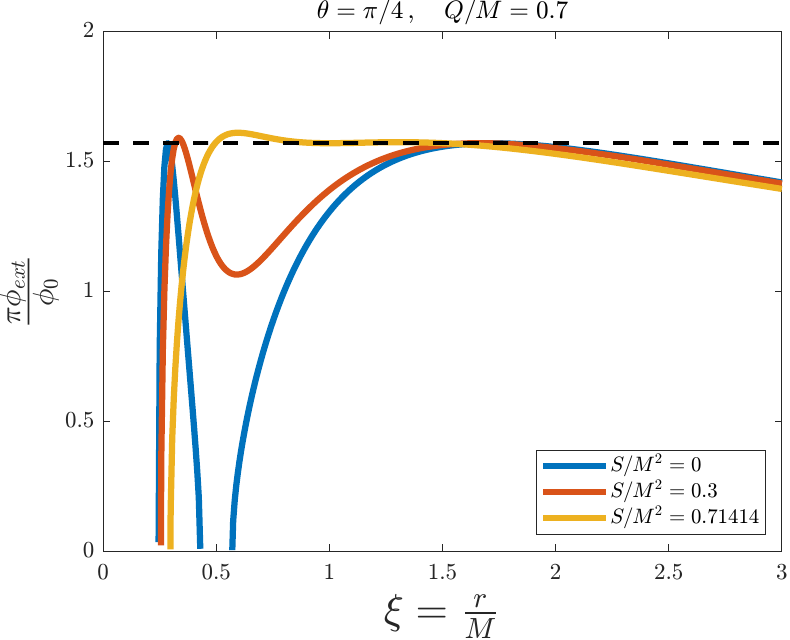}
\caption{External magnetic flux $\pi\phi_{\text{ext}}/\phi_0$ versus the adimensional distance $\xi=r/M$ for the simulation of a Kerr--Newman black hole with fixed $\theta=\pi/4$ and $A=Q/M=0.7$, and for different values of $B=S/M^2$. The dashed line represents the threshold value $\pi\phi_{\text{ext}}/\phi_0=\pi/2$.}
\label{fig:kn_pi4_0.7}
\end{figure}

In Figure \ref{fig:kn_pi4_0.7}, $\theta=\pi/4$ and $A=0.7$ are fixed. In this radial section of the spacetime, the ergosphere is now present, as can be seen in the figure when the magnetic flux exceeds its threshold value. The blue line represents the Reissner--Nordstr\o m limit, and the yellow one the extremal Kerr--Newman black hole, for this set of fixed parameters. As with \mbox{Figure \ref{fig:kn_0_0.7}}, the charge parameter $A$ is below its splitting value $1/\sqrt{2}$, so we can see a splitting of the flux until $B$ reaches a value between 0 and 0.3. Again, the simulation cannot reach values close to $r=0$, as is the case with every charged black hole. As we discussed for the simulations of Kerr black holes, the ergosphere cannot be reproduced with this setup. Therefore, this simulation is not appropriate for radial sections of rotating black holes away from their rotation axes.

\begin{figure}[H]
\includegraphics[width=10.5 cm]{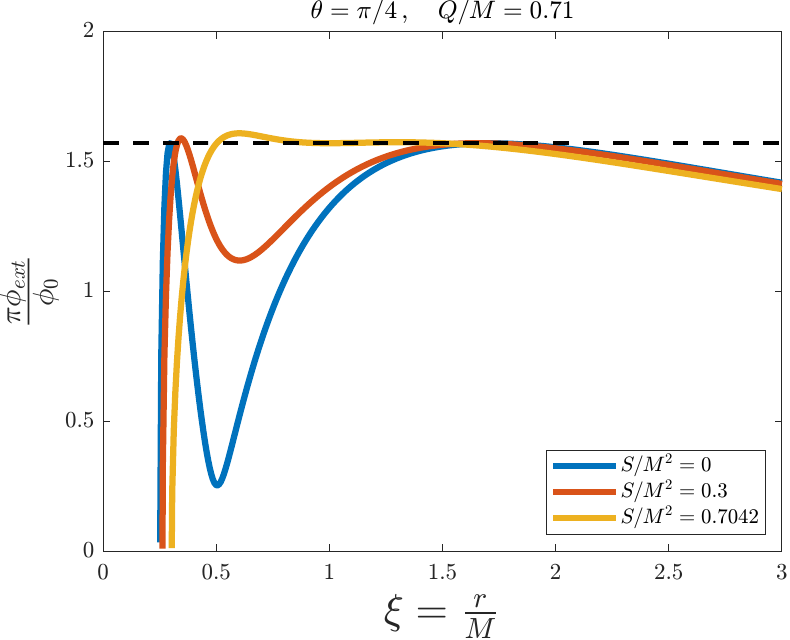}
\caption{External magnetic flux $\pi\phi_{\text{ext}}/\phi_0$ versus the adimensional distance $\xi=r/M$ for the simulation of a Kerr--Newman black hole with fixed $\theta=\pi/4$ and $A=Q/M=0.71$, and for different values of $B=S/M^2$. The dashed line represents the threshold value $\pi\phi_{\text{ext}}/\phi_0=\pi/2$.}
\label{fig:kn_pi4_0.71}
\end{figure}

In Figure \ref{fig:kn_pi4_0.71}, with $\theta=\pi/4$ and $A=0.71$, the same discussion applies as with Figure \ref{fig:kn_pi4_0.7}, except that the flux does not split for any value of $B$, as in Figure \ref{fig:kn_0_0.71}, because $B=0.71>1/\sqrt{2}$.

\vspace{-9pt}

\begin{figure}[H]
\includegraphics[width=10.5 cm]{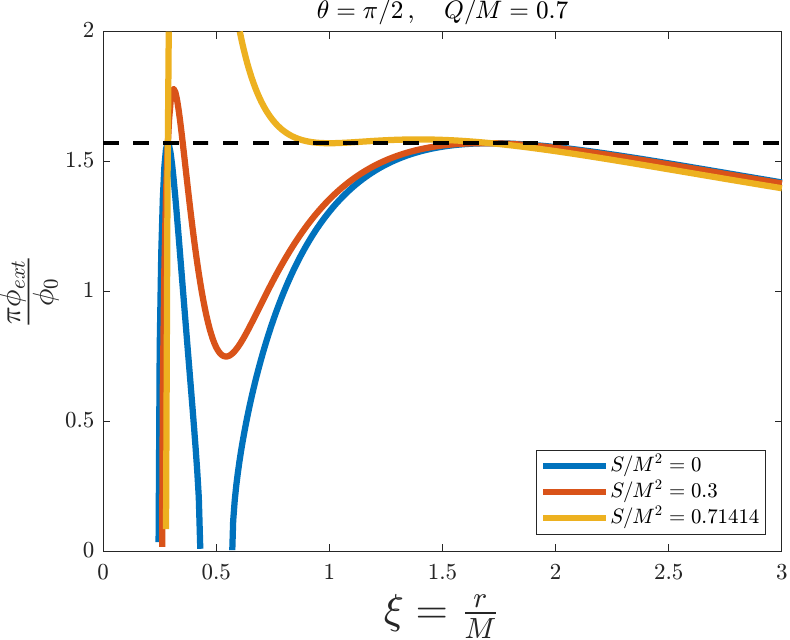}
\caption{External magnetic flux $\pi\phi_{\text{ext}}/\phi_0$ versus the adimensional distance $\xi=r/M$ for the simulation of a Kerr--Newman black hole with fixed $\theta=\pi/2$ and $A=Q/M=0.7$, and for different values of $B=S/M^2$. The dashed line represents the threshold value $\pi\phi_{\text{ext}}/\phi_0=\pi/2$.}
\label{fig:kn_pi2_0.7}
\end{figure}

\begin{figure}[H]
\includegraphics[width=10.5 cm]{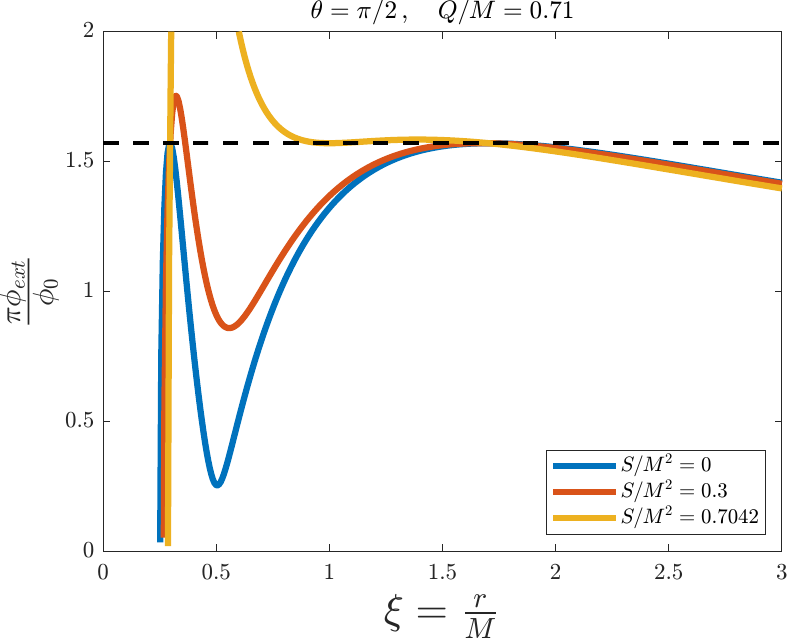}
\caption{External magnetic flux $\pi\phi_{\text{ext}}/\phi_0$ versus the adimensional distance $\xi=r/M$ for the simulation of a Kerr--Newman black hole with fixed $\theta=\pi/2$ and $A=Q/M=0.71$, and for different values of $B=S/M^2$. The dashed line represents the threshold value $\pi\phi_{\text{ext}}/\phi_0=\pi/2$.}
\label{fig:kn_pi2_0.71}
\end{figure}

Figures \ref{fig:kn_pi2_0.7} and \ref{fig:kn_pi2_0.71}, with $\theta=\pi/2$, and $B=0.7$ and $B=0.71$, respectively, show the same behaviours as Figures \ref{fig:kn_pi4_0.7} and \ref{fig:kn_pi4_0.71}, with even bigger non-simulable regions, as the radial size of the ergosphere is maximal at the equator of the black hole ($\theta=\pi/2$).

\vspace{-9pt}

\section{Summary and Conclusions}

In conclusion, following the procedure of~\cite{exotic_spacetimes_Sab_n_2018}, we have studied the possible simulations of radial sections of black hole spacetimes using a SQUID array embedded on an open transmission line, for Schwarzschild, Reissner--Nordstr\o m, Kerr and Kerr--Newman black holes. For the non-rotating black holes (Schwarzschild and Reissner--Nordstr\o m), the generation of an event horizon could, in principle, be achieved with this method. For this, we should avoid reaching magnetic flux values close to its critical value $\phi_0/2$ along a large region of the array---optimally reducing this region to a single SQUID, as discussed in~\cite{squid_wormhole_PhysRevD.94.081501,exotic_spacetimes_Sab_n_2018}. If the event horizon is successfully generated, Hawking radiation might be expected to appear in the system, due to quantum fluctuations of the electromagnetic vacuum at the horizon; but this can only be verified by carrying out the experiment and measuring radiation at both sides of the array with coincidence detection, as is proposed in~\cite{quantum_vacuum_hawking_radiation_RevModPhys.84.1}. 

Regarding rotating black holes (Kerr and Kerr--Newman), it seems impossible to simulate ergospheres  with our choice of radial sections of the spacetime. This may be overcome by choosing other types of 1~+~1D sections---for instance, orbital sections at $r$ constant, which could be an interesting follow up on this work. Other interesting extensions, such as the analysis of relevant observables, specific black-hole masses or higher dimensions, lie beyond the limited scope of the current work.




\vspace{6pt} 




\funding{\textls[-25]{C.S. has received financial support through the Ram\'on y Cajal programme (RYC2019-028014-I)}.}

\end{paracol}
\reftitle{References}

\end{document}